\documentclass[reprint,
nofootinbib,
 amsmath,amssymb,
 aps,
prb,
]{revtex4-2}

\usepackage{graphicx}
\usepackage{dcolumn}
\usepackage{bm}
\usepackage{physics}
\usepackage{hyperref}
\hypersetup{
    colorlinks=true,
    linkcolor=black,
    filecolor=black,      
    urlcolor=black,
    citecolor=blue,
    pdftitle={Overleaf Example},
    pdfpagemode=FullScreen,
    }
    
\usepackage{cleveref}
\usepackage{xcolor}
\usepackage{bbold}
\usepackage{lipsum}
\usepackage[normalem]{ulem}
\allowdisplaybreaks

\usepackage{amsmath}
\usepackage{mathtools}
\usepackage{tikz}

\usetikzlibrary{decorations.markings,positioning}
\usetikzlibrary{shapes.misc}
\tikzset{cross/.style={cross out, draw=black, minimum size=2*(#1-\pgflinewidth), inner sep=0pt, outer sep=0pt},
cross/.default={1pt}}

\begin{document}

\preprint{APS/123-QED}

\title{Quantum state preparation and readout with modulated electrons}

\author{J. Abad-Arredondo$^{1,2}$}
\email{jaime.abad(at)uam.es}
\affiliation{$^{1}$Departamento de F\'isica Te\'orica de la Materia
Condensada, Universidad Aut\'onoma de Madrid, E- 28049 Madrid, Spain} 
\affiliation{$^{2}$Condensed Matter Physics Center (IFIMAC),
Universidad Aut\'onoma de Madrid, E- 28049 Madrid, Spain}
\author{A. I. Fern\'andez-Dom\'inguez$^{1,2}$}%
\affiliation{$^{1}$Departamento de F\'isica Te\'orica de la Materia
Condensada, Universidad Aut\'onoma de Madrid, E- 28049 Madrid, Spain} 
\affiliation{$^{2}$Condensed Matter Physics Center (IFIMAC),
Universidad Aut\'onoma de Madrid, E- 28049 Madrid, Spain}

\date{\today}

\begin{abstract}
We provide a comprehensive study of the capabilities of modulated electron wavefunctions for the preparation and readout of the quantum state of the quantum emitters (QEs) they interact with. First, we consider perfectly periodic electron combs, which do not produce QE-electron entanglement, preserving the purity of the QE while inducing Rabi-like dynamics in it. We extend our findings to realistic, non-ideally modulated electron wavepackets, showing that the phenomenology persists, and exploring their use to prepare the emitter in a desired quantum state. Thus, we establish the balance that electron comb size, emitter radiative decay, and electron-emitter coupling strength must fulfil in order to implement our ideas in  experimentally feasible platforms. Finally, moving into the limit of small electron combs, we reveal that these wavefunctions allow for quantum state tomography of their target, providing access not only to the populations, but also the coherences of the QE density matrix. We believe that our theoretical results showcase modulated free-electrons as very promising tools for quantum technologies based on light-matter coupling. 

\end{abstract}

\maketitle

\section{Introduction}
The ability to control the state of a given qubit plays a central role in all the emergent branches of quantum technologies. Different applications such as quantum computing, communication or metrology benefit from having precise control over the state of quantum systems. Proposals of the exploitation of optical fields to prepare and manipulate quantum states date back to the early stages of quantum mechanics itself \cite{Rabi1937}, when both quantum and semi-classical formulations showed that optical pulses and oscillatory fields provide a useful toolkit for such purpose. Despite the undeniable success in the use of light fields in quantum technologies, free space optical excitations suffer from being diffraction-limited, which imposes a strong limitation on the minimal size that elements must present in order to be individually addressable. 

On the other hand, current state-of-the-art electronic microscopes allow sub-nanometric resolution and high monocromaticity, with electron energies spreading down to the meV level~\cite{Erni2009,Morishita2016,Krivanek2014}. These techniques use the evanescent character of free electron fields~\cite{Jackson1999} to explore photonic and material excitations with a spatial selectivity that is innately larger than that attainable in optical spectroscopy. This has been exploited to probe quantum features in different targets~\cite{GarciaDeAbajo2010,DiGiulio2019,Yankovich2019,Polman2019}. The advent of photon-induced near-field electron microscopy (PINEM)~\cite{Barwick2009} has shown that by making free electrons interact with a driven optical resonator their wavevector distribution becomes discrete (as a consequence of the exchange of individual photons with the target). The drift of these \emph{modulated electrons}, which present a structured wavefunction in wavevector space, after the PINEM interaction reshape their spatial wavefunction into trains of tightly focused subbunches. This strategy has allowed the coherent control of the wavefunction of a free electron~\cite{Barwick2009,Zewail2010,Feist2015}, as well as the preparation of ultrashort electron pulses, which makes the probing of ultrafast dynamics with nanometric resolution possible ~\cite{Barwick2009,Breuer2013,Priebe2017,Kozak2017,Kozak2018,Morimoto2018}. This has led to striking results, such as the recording of the charge dynamics in out-of-equilibrium material systems~\cite{Yannai2023}, or the sub-optical cycle dynamics of electromagnetic fields~\cite{Nabben2023}, and their associated phase distribution~\cite{Gaida2024,Bucher2024}.

Recent advances in the shaping of the free-electron wavefunctions have opened the way towards the realization of quantum optics with electron beams~\cite{Zewail2010,Talebi2018,Kfir2020,Kaminer2020,Reinhardt2020,Kfir2019}. Thus, the study of quantum coherence in the interaction of free electrons with both extended, optical fields~\cite{DiGiulio2020,Kfir2021,Baranes2022} and localized, exciton transitions in quantum emitters (QEs),~\cite{AbadArredondo2024,Zhao2021,Zhang2021,Ruimy2021,Ratzel2021,Morimoto2021,Zhang2022,Ran2022,GarciadeAbajo2022} has experienced a surge. Most works have focused on high energy beams that operate in the so-called {\it non-recoil approximation}, which enable the direct analysis of the electron-target interaction itself. Theoretical investigations modeling realistic systems~\cite{Zhang2021,Ran2022,Morimoto2021,Ratzel2021} indicate that electronic bunching is key for the wavefunction engineering, as it leads to measurable changes in the excitation probability of quantum targets, such as QEs. Other reports have demonstrated entanglement generation in emitter ensembles using free electrons, and the possibility of inducing Rabi dynamics in them~\cite{Zhao2021,Ratzel2021}. These results point at the prospect of using free electrons to prepare and characterize the quantum state of QEs~\cite{Ruimy2021,Zhang2022}.

The theoretical framework for the quantum description of the coherent interaction between free electrons and arbitrary optical and material excitations is Macroscopic QED~\cite{Buhmann2012,Feist2020}, which enables the construction of interaction Hamiltonian models in terms of the electromagnetic Dyadic Green's function~\cite{AbadArredondo2024}. In this paper, we tackle the general problem of the assessment of free electrons as tools for QE state generation and interrogation. We therefore parameterize our models in a phenomenological fashion (within realistic bounds), avoiding the  question of their actual physical implementation. We consider first a single electron-QE interaction event, and identify the self-correlation of the electron wavefunction in wavevector space, rather than spatial bunching, as the relevant figure of merit in this configuration. Next, in the limit of perfect modulation, we find that the QE purity is maintained during the interaction and free electrons and emitter remain disentangled. We focus next on the phenomenology that arises under  continuous driving with modulated free electrons. We show that electron-induced dephasing and the QE radiative decay compete with Rabi oscillations in the quantum dynamics of the system, which enables us to discern the attainable steady states. Finally, we focus on free-electron-assisted state tomography, and determine that small electrons are best suited for the task. We then propose a two-electron measurement protocol that is able to determine QE state purity.

The paper is structured as follows: In Section II, we introduce the Hamiltonian describing the interaction between modulated electrons and single QEs. Next, we explore perfectly modulated electrons in Sec. III, and show how single electron interaction and continuous electron pumping can be used to perform QE state preparation. We then proceed to analyze realistic, finitely modulated electronic wavefunctions and QE spontaneous emission to set the constraints that the system parameters must fulfil for the implementation of targeted QE states in Sec. IV. Finally, in Sec. V, we outline the potential of free electrons, and their quantum nature, as tools for quantum state tomography.

\section{Theory and model system}\label{sec:theory}
The interaction of a localized material transition and a passing electron can be described through the Hamiltonian \cite{AbadArredondo2024,Zhao2021}:
\begin{gather}
    \hat{H}_I=\hbar \sum_q [g_q\hat{\sigma}^\dagger \hat{b}_q + g_q^*\hat{\sigma}\hat{b}_q^\dagger], \label{eq:IntHam0}
\end{gather}
where $g$ is the coupling strength between the free electrons and the QE\footnote{Equivalent analytical expressions for the case of a dipolar transition are given in several works~\cite{AbadArredondo2024,Zhao2021,Ruimy2021,GarciadeAbajo2022}.}. $\hat{\sigma}=\ket{g}\bra{e}$ is the usual two-level-system annihilation operator modelling the QE exciton, and $\hat{b}_q=\sum_k \ket{k-q}\bra{k}$ is a ladder operator that shifts the free-electron wavefunction in $k$-space by a wavevector $q$. Note that we assume a reciprocal space discretization in a box of arbitrary length $L$~\cite{Zhao2021,AbadArredondo2024}. To analyze the effect of the passing electron on the QE, we calculate the final state of the system using the scattering matrix formalism, by integrating the propagator of the Schrödinger Equation,
\begin{gather}
    \ket{\psi}_f=\mathcal{T}\exp\left[-\frac{i}{\hbar}\int_{-\infty}^{\infty} dt \, \hat{H}_{I,{\rm int}}(t)\right]\ket{\psi}_i\equiv \hat{S}\ket{\psi}_i, \label{eq:TimeOrdering_scatteringmatrix}
\end{gather}
where $\hat{H}_{I,{\rm int}}(t)$ is the interaction Hamiltonian in the interaction picture and $\mathcal{T}$ is the time ordering operator. 

By use of the Magnus expansion~\cite{Blanes2010}, a propagator with true exponential form can be obtained using a series expansion. Using the non-recoil approximation and the energy dispersion of free electrons, the propagator obtained from the Magnus expansion up to second order (see Appendix \ref{sec:Magnus_expansion_first_2_terms}) is given by
\begin{gather}
    \hat{S}=\exp\Bigg[-i \left( \beta \hat{\sigma}^\dagger \hat{b} +\beta^* \hat{\sigma} \hat{b}^\dagger \right)+ \hat{O}(\beta^3)\Bigg],\label{Magnus}
\end{gather}
where we have introduced $\beta \equiv g_{q} L /v_0$ as the integrated interaction strength ($v_0$ is the central velocity of the electron beam). Note that we have dropped the $q$ subindex as $q=\omega/v_0$ from now on ($\omega$ is the natural frequency of the QE). Using that $(\beta\hat{\sigma}^\dagger \hat{b} + \beta^*\hat{\sigma}\hat{b}^\dagger)^2=\left|\beta\right|^2 \mathbb{1}$, and absorbing complex phase of $\beta$ in the QE ground state, the scattering matrix can be written as~\cite{Ruimy2021,Zhang2022}
\begin{equation}
    \hat{S}=\mathbb{1}\cos(\left|\beta\right|)-i(\hat{\sigma}^\dagger \hat{b}  + \hat{\sigma}\hat{b}^\dagger)\sin(\left|\beta\right|) \label{eq:Propagator_simp}.
\end{equation}
This simple expression allows describing the free-electron-QE interaction, keeping in mind that corrections of order $\beta^3$ would need to be included in Equation~\ref{Magnus} for large coupling strengths. In what follows, we omit the absolute value of the coupling strength, $\beta$, and assume that it is real and positive. For QE state preparation and readout, the initial state of the compound system can be written as the product state of QE and free-electron initial matrix densities, $\rho_0=\rho_0^{\rm QE}\otimes \rho_0^e$, and the final state after interaction is simply given by $\rho=\hat{S}\rho_0\hat{S}^\dagger$. In the next sections, we employ the closed-form approach outlined here to investigate different free-electron and QE configurations. 

\section{\label{section:Ideally_modulated_electrons} Ideally modulated electrons}  Similarly to coherent optical states, ideally modulated electron wavefunctions are the eigenstate of the ladder operator $\hat{b}$. A particular state that fulfills this condition can be written as~\cite{Baranes2023}
\begin{gather}
    \ket{C_\phi}=A \sum_{n=-\infty}^{\infty} e^{i n \phi} \ket{k_0 -nq}, \label{eq:ideally_mod_electron_wavefunction}
\end{gather}
with $A$ being a normalization factor. Note that Equation~\eqref{eq:ideally_mod_electron_wavefunction} is not unique, and particle-like wavefunctions can also be used to describe ideally modulated electrons~\cite{Zhang2021,Zhang2022,Ran2022}. The action of the ladder operator on this state is just $\hat{b}\ket{C_\phi}=e^{-i\phi}\ket{C_\phi}$, and conversely $\hat{b}^\dagger\ket{C_\phi}=e^{i\phi}\ket{C_\phi}$. 

The action of the scattering matrix on an initial state characterized by the product of this electronic wavefunction and any arbitrary QE state, $\rho^{\rm QE}_0$, has the form 
\begin{gather}
    \rho=\left[\hat{S_r} \rho^{\rm QE}_0\hat{S_r}^\dagger \right] \otimes \dyad{C_\phi},
\end{gather}
with $\hat{S_r}=\mathbb{1}\cos(\beta)-i(\hat{\sigma}^\dagger e^{-i\phi} + \hat{\sigma} e^{i\phi})\sin(\beta)$. Thus, the final state turns out to be a product state of the initial electronic comb and an unitary operation applied onto the inial QE state. Hence, free electron and QE remain disentangled, and the passing electron does not retain any information about the emitter. This indicates that ideally modulated electrons will perform poorly as tools for quantum state readout but, as we show below, has other interesting consequence. By tracing out the free-electron degrees of freedom in the final density matrix, one finds $\rho^{\rm QE}={\rm Tr}_e[\rho]=\hat{S_r}\rho^{\rm QE}_0\hat{S_r}^\dagger$. The unitary character of $\hat{S_r}$ then yields ${\rm Tr}\{(\rho^{\rm QE})^2\}={\rm Tr}\{(\rho^{\rm QE}_0)^2\}$. This implies that the purity of the QE state, 
\begin{gather}
\mathcal{P}=\sqrt{2 \Tr{(\rho^{\rm QE})^2}-1},\label{eq:Purity definition}
\end{gather}
is preserved in its interaction with a perfectly modulated electron beam regardless of $\beta$, which is a desirable property for quantum state preparation. Note that $\mathcal{P}=1$ ($\mathcal{P}=0$) for pure (maximally mixed) QE states in Equation~\eqref{eq:Purity definition}.  

We consider now that the QE is initially in the pure state $\ket{\psi^{\rm QE}}=\cos(\theta)\ket{g}+\sin(\theta)e^{-i\gamma}\ket{e}$ and the electron is ideally modulated with wavefunction given by Equation~\eqref{eq:ideally_mod_electron_wavefunction}. It can be shown that the probability of finding the QE in the ground state after interaction (i.e. the final ground state population) is given by
\begin{flalign}
\expval{\rho^{\rm QE}}{g}_{\tiny{\mbox{ideal}}}=&\cos^2(\theta)-\sin^2(\beta)\cos(2\theta) \nonumber\\
&+\frac{1}{2}\sin(\phi-\gamma)\sin(2\theta)\sin(2\beta).\label{eq:Probability_gs_ideally_moduated}
\end{flalign}
Conversely, if the initial state of the electron is monochromatic (\emph{mc}), i.e equal to $\ket{k_0}$, then the same probability becomes $\expval{\rho^{\rm QE}}{g}_{\tiny{mc}}=\cos^2(\theta)-\sin^2(\beta)\cos(2\theta)$. Hence, we can identify the second line of Equation~\eqref{eq:Probability_gs_ideally_moduated} as the contribution due to electron modulation. This term presents a dependence on $\sin({2\theta})$, which can be related to the coherences of the $\rho_0^{\rm QE}$, and indicates that the contribution to the final populations takes place through interference with the initial state.  The modulation contribution is maximal when $|\sin(2\theta)|=1$, a  condition that coincides with the maximum expectation value for the QE dipole moment operator, $\hat{\mathbf{d}}\equiv \mathbf{d} \hat{\sigma} + \mathbf{d}^*\hat{\sigma}^\dagger$. Interestingly, the term is also sensitive to the initial phase, $\gamma$, through $\phi$. This indicates that through modulation one may have access to complete information of the QE state. 

Although large coupling strengths are attainable under sufficiently precise electron positioning~\cite{AbadArredondo2024}
and therefore single electron-QE interaction events are already of practical relevance, in general, the coupling between a QE and a single free electron will be weak. It is therefore of interest to investigate setups in which the QE sequentially interacts with identically prepared modulated electrons. Assuming that these arrive at the emitter with a constant rate $\gamma_e=1/\tau$ (much slower than the QE-electron interaction), and using that each of them alter the target state by $\Delta\rho^{\rm QE}=\hat{S_r} \rho^{\rm QE}\hat{S_r}^\dagger- \rho^{\rm QE}_0$, we can construct a Von Neumann equation for the QE dynamics. In the limit of weak coupling, ${\hat{S_r}\rightarrow\mathbb{1}-i\beta(\hat{\sigma}^\dagger e^{-i\phi} + \hat{\sigma} e^{i\phi})}$ and
\begin{flalign}
    \dot{\rho^{\rm QE}}\approx\frac{\Delta\rho^{\rm QE}}{\tau}
    \approx -\frac{i}{\hbar}[\hat{H}_{\rm eff},\rho^{\rm QE}],\label{Rabi_H}
\end{flalign}
where $\hat{H}_{\rm eff}= \frac{\hbar}{\tau}\beta(\hat{\sigma}^\dagger e^{-i\phi} + \hat{\sigma} e^{i\phi})
$. This effective Hamiltonian has the form of a coherent optical driving of the QE in the rotating frame~\cite{Scullybook}, and therefore we can expect the emergence of Rabi oscillations in the interaction with the train of modulated electrons.

We now introduce the Pauli matrices $\hat{\sigma}_i$, defined as $\hat{\sigma}_1\equiv \dyad{e}{g}+\dyad{g}{e}$, $\hat{\sigma}_2\equiv -i\dyad{e}{g}+i\dyad{g}{e}$ and $\hat{\sigma}_3\equiv \dyad{e}{e}-\dyad{g}{g}$, which enable us to parametrize the QE state as $\rho^{\rm QE} = (\mathbb{1}+x\hat{\sigma}_1+y\hat{\sigma}_2+z\hat{\sigma}_3)/2$ in the Bloch sphere, where the 2 eigenstates of $\hat{H}_{\rm eff}$ above have coordinates $\pm[\cos(\phi),\sin(\phi),0]$. The component of any QE state along these eigenvectors is preserved over time, while its normal component oscillates. Thus, the effective Hamiltonian above induces a state precession around the axis defined by $\phi$, the relative phase between consecutive terms in Equation~\eqref{eq:ideally_mod_electron_wavefunction}. As this vector is normal to the $z$-axis, a QE initially in the ground state (or, in general, any state normal to the eigenvectors of $H_{\rm eff}$), will experience Rabi-like oscillations in its dynamics, with frequency $\beta\gamma_e$, and the phase of the coherences in $\rho^{\rm QE}$ fixed in time. Note that, by adiabatically modifying the modulation phase $\phi$ of the incoming electrons, it is possible to rotate the phase of the coherences in $\rho^{\rm QE}$. This, together with the Rabi dynamics above, constitutes a complete mechanism for arbitrary QE state preparation.

\section{State preparation with realistic electron wavefunctions}\label{sec:stateprep}
We have shown that perfectly modulated electrons present promising features for the preparation and readout of quantum states in QE targets. In this section, we explore the constraints for this purpose that emerge when non-ideal, realistic electron modulations are considered.

\subsection{Single electron-QE interaction}
We inspect first electron wavefunctions spanning a continuum of wavevector components, with the general form $\ket{\psi^e_0}=\int dk B(k)\ket{k}$. Thus, the initial state of the QE-electron system is $\rho_0=\rho^{\rm QE}_0\otimes\dyad{\psi^e_0}$. Applying  Equation~\eqref{eq:Propagator_simp} and taking the partial trace, ${\rho^{\rm QE}=\int dk \bra{k}\hat{S}\rho_0 \hat{S}^\dagger\ket{k}}$, we obtain
\begin{widetext}
\begin{flalign}
\rho^{\rm QE}=&\rho^{\rm QE}_0+ i   \frac{\sin(2\beta)}{2} \left[\rho^{\rm QE}_0, I_1^{'}\hat{\sigma}_1+I_1^{''}\hat{\sigma}_2\right] \nonumber \\
&+\frac{\sin^2(\beta)}{2} \left\{\left(\hat{\sigma}_1 \rho^{\rm QE}_0 \hat{\sigma}_1+\hat{\sigma}_2\rho^{\rm QE}_0 \hat{\sigma}_2-2\rho^{\rm QE}_0\right)+I_2^{'}\left(\hat{\sigma}_1\rho^{\rm QE}_0\hat{\sigma}_1-\hat{\sigma}_2\rho^{\rm QE}_0\hat{\sigma}_2 \right)+ I_2^{''}\left(\hat{\sigma}_1\rho^{\rm QE}_0\hat{\sigma}_2+\hat{\sigma}_2\rho^{\rm QE}_0\hat{\sigma}_1 \right)\right\},\label{eq:final_dens_mat_QE_c}
\end{flalign}
\end{widetext}
 where the effect of the initial wavefunction is encoded in the overlapping integrals $I_1\equiv\expval{\hat{b}}_e=\int dk\,  B(k) B^*(k+q)$ and $I_2\equiv\expval{\hat{b}^2}_e=\int dk\,  B(k) B^*(k+2q)$. The prime and double prime notation is used to indicate the real and imaginary part of these magnitudes, respectively. One can see that when $B(k)$ is composed of a single wavevector component narrower than $q=\omega/v_0$, these integrals vanish, while $I_1=I_2=1$ for a perfectly modulated wavefunction like Equation~\eqref{eq:ideally_mod_electron_wavefunction}. Thus, $I_1$ and $I_2$ account for the impact that the free-electron modulation has on the final QE state. 

Equation~\eqref{eq:final_dens_mat_QE_c} is completely agnostic to  the choice of $\ket{\psi^e_0}$, and allows the comparison among different wavefunction configurations at all interaction orders within the validity of the Magnus expansion in Equation~\eqref{Magnus}. In Ref.~\cite{Zhang2021}, gaussian-shaped $B(k)$ were investigated, for which $I_{1,2}\rightarrow1$  in the particle-like limit of diverging width (in $k$-space). This translates into a large impact of electron modulation in $\rho^{\rm QE}$, while it is irrelevant in the wave-like, unmodulated limit of vanishing gaussian width (for which $I_{1,2}=0$). A different wavefunction shaping strategy, based on PINEM, has been  investigated recently~\cite{Feist2015,DiGiulio2020}, as it enables the preparation of free electrons with a comb-like wavevector distribution of the form $\ket{\psi_e}=\sum_l J_l(2\beta)\ket{k_0 +l q}$ (where $J_l$ is the l-th order Bessel function). It can be shown that $I_n=\delta_{n,0}$ ($n=1,2$) in this case, which means that the instantaneous modulation induced by the interaction with a classical optical field is useless. However, if the free electrons are left to drift after the PINEM interaction~\cite{Morimoto2021,Zhang2022}, contributions to the overlapping integrals appear for relativistic velocities. In Appendix \ref{sec:Mod_integrals_Gover}, we explore this mechanism and, in agreement with Ref.~\cite{Zhao2021}, we find that $|I_1|\lessapprox 0.581$ in this configuration. The difference between drifted-PINEM-like and particle-like electron wavepackets becomes more apparent when considering targets in which different transition energies are present. $I_{1,2}$ have the form of autocorrelation functions in $k$-space. A particle-like electron has a very broad wavevector distribution, and therefore it is self-similar upon any wavevector translation, interacting in the same form to all the optical transitions in the target. On the other hand, the comb-shaped wavefunction of PINEM electrons allow to address selectively certain transitions of the target~\cite{AbadArredondo2024}. This means that the effect of electron modulation must be understood as a result of quantum interference, rather than classical charge localization or bunching~\cite{Zhang2021}, and that strategies employing quantum-optical fields~\cite{DiGiulio2019} or multiple resonant frequencies~\cite{Priebe2017,Reinhardt2020} are required to fully exploit modulated electron wavepackets for QE state preparation.  


Back to Equation~\eqref{eq:final_dens_mat_QE_c}, its first line contains the term that gives rise to the Rabi dynamics discussed in Section~\ref{section:Ideally_modulated_electrons} (note that $\sin(2\beta)/2\approx\beta$ in the weak coupling limit). Importantly, it shows that the Rabi frequency is proportional to $|I_1|$, and therefore originates from the electron modulation. Moreover, this is the leading order term in $\beta$ (note that $\sin^2(\beta)\approx\beta^2$ in the second line~\cite{Zhao2021}), which reveals that the phenomenology for realistic electron modulation resembles the ideal one for sufficiently weak coupling strength. The last two terms in Equation~\eqref{eq:final_dens_mat_QE_c} also present a dependence on the electron modulation through $I_2$. As we will later show, these contributions are responsible for cancelling the electron-induced dephasing that takes place for non-modulated electrons at large coupling strengths. 

\begin{figure}[t!]
\includegraphics[angle=0,width=\columnwidth]{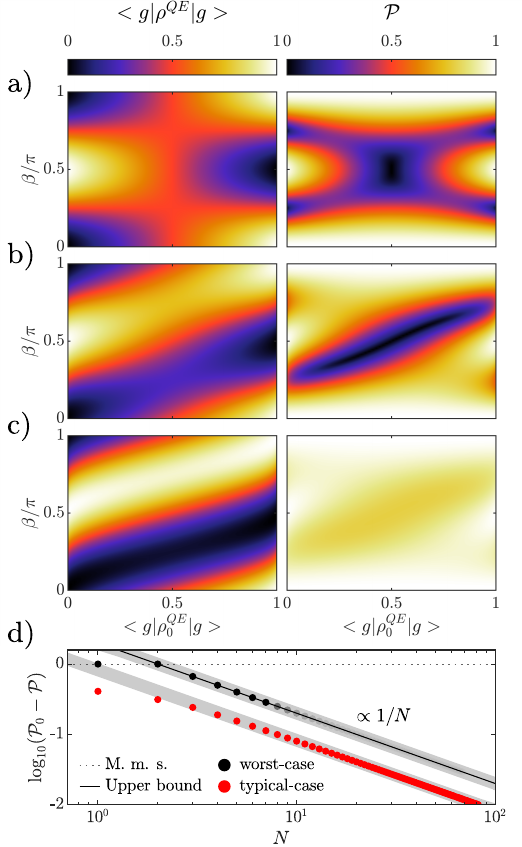}
\caption{Final state of the QE after its interaction with a realisitic (non-ideally modulated) electron wavefunction. The initial QE state is $\ket{\psi_{0}^{QE}}=\cos(\theta)\ket{g}+\sin(\theta)e^{i\frac{\pi}{2}}\ket{e}$. The left (right) panels in (a)-(c) display the ground state probability (purity) for the final QE state. The comb sizes are $N=1$ (a), 2 (b) and 10 (c). Panel (d) renders the purity loss, $\Delta\mathcal{P}=\mathcal{P}_0-\mathcal{P}$ as a function of the electron comb size, $N$.} \label{fig:fig1}
\end{figure}

To illustrate the difference in quantum state preparation between realistic and perfect electron modulation, we choose a free-electron wavefunction consisting of a finite comb of $N$ identical peaks, $\ket{\psi^e_0}=\tfrac{1}{\sqrt{N}}\sum_{m=1}^N \ket{k-mq}$, for which $I_n=1-n/N$ ($I_2=0$ for $N=1$). It interacts with a QE that is initially in the pure state $\ket{\psi_{0}^{\rm QE}}=\cos(\theta)\ket{g}+\sin(\theta)e^{i\frac{\pi}{2}}\ket{e}$. Figure~\ref{fig:fig1} (a-c) show the probability of finding the QE in the ground state after the interaction (left) and the purity (right) of its final state as a function of the initial ground state population, $\expval{\rho^{\rm QE}}{g}=\cos^2(\theta)$, and the coupling strength, $\beta/\pi$. Results for combs of three different sizes, $N=1$ (a), 2 (b) and 10 (c) are displayed. In the case of a monochromatic electron ($N=1$), we can observe two different regions in the dependence of the ground state population on $\beta$. At very low and very high couplings, the final ground state population follows the initial one, while the trend is the opposite at intermediate $\beta$, which indicates that the effect of the interaction with the free electrons is the largest~\cite{Zhang2022, GarciadeAbajo2022}. Note that when the initial ground state probability is $0.5$, the ground state probability after interaction remains unaltered and only purity loss takes place, with the extreme case of vanishing $\mathcal{P}$ in the region $\expval{\rho^{\rm QE}}{g}\approx\beta/\pi\approx0.5$, which reveals that the QE is left in a maximally mixed state. This is detrimental for quantum state preparation, but potentially beneficial for the readout of the QE state, which can become highly entangled with the passing electrons (see Section \ref{section:Quantum_state_tomography}). Note that the purity of the final state approaches unity in most QE and electron wavefunction configurations apart from this central region of the map.  

The $N=1$ configuration represents the opposite of the ideally modulated electron treated in Section~\ref{section:Ideally_modulated_electrons}. As the number of wavevector components of the electron comb increases, the purity loss is reduced in the whole parameter plane, as shown in Figure \ref{fig:fig1}(b-c). This trend is accompanied by a modification of the landscape of final ground state probabilities, which develops a more complex dependence on $\ket{\psi^e_0}$ and coupling strength, approaching Equation~\eqref{eq:Probability_gs_ideally_moduated} as $N$ increases. This allows, for instance, to prepare the QE in the excited state with a wider range of system parameters~\cite{GarciadeAbajo2022}. The maps show that the purity loss is larger for QEs initially in a equally weighted superposition of ground and excited state for all $N$. Like in the perfectly modulated case, this can be linked to the maximum expectation value of the QE dipole moment,  a condition in which the coherences in $\rho_0^{\rm QE}$ acquire a relevant role in the QE-electron interaction (see Section~\ref{section:Ideally_modulated_electrons}). In this particular configuration, an upper bound for the QE purity loss can be extracted, having
\begin{gather}
    \Delta\mathcal{P}\leq
    1-\frac{1}{N}\sqrt{(N-2\sin^2(\beta))^2 +\sin^2(2\beta)}\rightarrow\frac{2\sin^2(\beta)}{N}, \label{eq:Purity_loss_upper_bound}
\end{gather}
where, in the last step, the limit $N\rightarrow \infty$ is taken. Equation~\eqref{eq:Purity_loss_upper_bound} reveals the impact of finite size effects in the electron comb on the purity of the QE state. It also unveils that $\Delta\mathcal{P}$ scales as $\beta^2$ in the weak coupling limit, vanishing faster than the Rabi frequency in the QE dynamics (proportional to $\beta$ in Equation~\eqref{Rabi_H}). 

In Figure~\ref{fig:fig1}(d) the purity loss of the QE state is plotted against the number of wavevector components in the electron wavefunction. The worst (largest loss) and typical (averaged) $\Delta\mathcal{P}$ are shown in black and red dots, respectively. They are extracted numerically from calculations for all possible values of $\{\beta,\theta\}$ and for $N$ ranging from 1 to $1000$. The black solid line renders the upper bound given by Equation~\eqref{eq:Purity_loss_upper_bound}, which overlaps with the worst-case purity for $N>1$. The faint gray thick lines are a guide for the eye of the $N^{-1}$ dependence featured by both sets of data for large enough electron comb. The dotted horizontal line corresponds to $\Delta\mathcal{P}=1$, for which the QE is in a maximally mixed state (M. m. s., $\mathcal{P}=0$) and can be maximally entangled with the passing electron. Notice that, unexpectedly, this condition can be met for $N>1$, which opens the door to exploiting modulated electrons for quantum state tomography.

\subsection{Continuous electron-QE interaction}

We study next the phenomenology of continuous electron-QE interaction for realistic wavepackets. Following the same arguments that led to Equation~\eqref{Rabi_H}, now fed with the density matrix in Equation~\eqref{eq:final_dens_mat_QE_c}, a master equation for the QE dynamics induced by its continuous, sequential interaction with non-ideally modulated electron wavepackets (with rate $\gamma_e$) can be obtained~\cite{Zhao2021}. Expressing $\rho^{\rm QE}$ in terms of the coordinates of the Bloch sphere, it can be casted as 
\begin{flalign}
\begin{pmatrix}
\dot{x} \\
\dot{y} \\
\dot{z}
\end{pmatrix} =&  \begin{pmatrix}
g_1I_2^{'}& g_1 I_2^{''} & g_2  I_1^{''}\\
g_1 I_2^{''}& -g_1I_2^{'} & - g_2 I_1^{'}\\
-g_2 I_1^{''}& g_2 I_1^{'}& -g_1
\end{pmatrix}
\begin{pmatrix}
x \\
y \\
z
\end{pmatrix} \nonumber \\
&-(\gamma_0+g_1) \mathbb{1}\begin{pmatrix}
x \\
y \\
z
\end{pmatrix}
-\begin{pmatrix}
0 \\
0 \\
\gamma_0
\end{pmatrix},
\label{eq:Master_eq_Bloch}
\end{flalign}
where, for compactness, we have defined $g_1=\gamma_e \sin^2(\beta)$ and $g_2=\gamma_e \sin(2\beta)$, and $I_{1,2}$ are the $k$-space overlapping integrals defined in Section~\ref{section:Ideally_modulated_electrons}. Note that, in a refinement of our model, we have introduced a Lindblad superoperator in the master equation describing the spontaneous, radiative decay of the QE, with rate $\gamma_0$.   

To shed light into the dynamics described by Equation~\eqref{eq:Master_eq_Bloch}, we examine first its steady state ($\dot{x}=\dot{y}=\dot{z}=0$), and subsequently analyze how it is reached. In absence of modulation ($I_{1,2}=0$, i.e. for a monochromatic electron), the QE steady state is
\begin{flalign}
     \begin{pmatrix} x_{\rm ss}\\ y_{\rm ss}\\ z_{\rm ss}\end{pmatrix}=-\begin{pmatrix} 0\\ 0\\ \frac{\gamma_0}{\gamma_0+2\gamma_e \sin^2(\beta)},\end{pmatrix} \label{ssmatrix},
\end{flalign}
 which nicely showcases the interplay between the two mechanisms governing the dynamics: radiative decay tends to bring the QE to it's ground state ($z=-1$ in the Bloch sphere), while the incoming electrons tend to drive the QE to a maximally mixed state at the origin of the Bloch sphere. Evaluating the probability of finding the QE in its excited state, $\expval{e}{\rho^{\rm QE}}=\tfrac{1+z_{\rm ss}}{2}$, we find that its maximum value is $1/2$~\cite{Zhao2021,Ratzel2021,Beer2024,GarciadeAbajo2022}, obtained for $\gamma_e\gg\gamma_0$. This condition yields $\mathcal{P}=z_{\rm ss}^2=0$, which reveals that all the quantum coherence in the QE is erased by the incoming electrons. This result is the extension of the purity loss discussed above for single electrons: In each interaction event, the QE state experiences a purity loss, and once the QE state is a statistical mixture, the steady state correspods to equal ground and excited state populations (i.e. when the QE is in a maximally mixed state). We term this mechanism {\it electron-induced dephasing}, which also plays a role in the QE dynamics once modulation is introduced in the free-electron wavefunction, although its steady state becomes more complex. 

\begin{figure}[t]
\includegraphics[angle=0,width=\columnwidth]{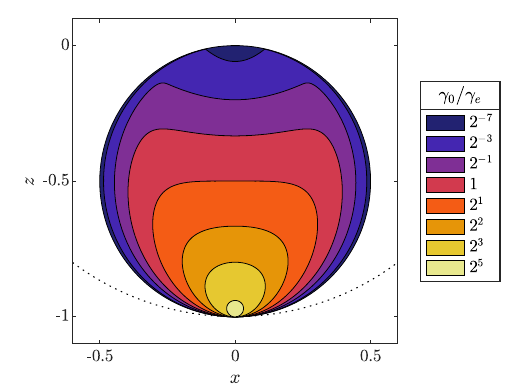}
\caption{QE steady states accessible through the continuous interaction with modulated electrons. Each colour corresponds to a value of $\gamma_0/\gamma_e$, the ratio between QE radiative decay and electron driving rates (the rest of parameters are free). Each region contains all the states attainable for larger $\gamma_0/\gamma_e$.} \label{fig:fig2}
\end{figure}

In Figure~\ref{fig:fig2}, we show the steady states accessible through electron modulation for values of the ratio $\gamma_0/\gamma_e$ ranging 3 orders of magnitude. Each color region contains the steady states accessible through variations in the overlapping integrals and coupling strength. The phases of the integrals are set so that the QE states lay within the $xz$-plane, but the regions acquire azimuthal symmetry around the $z$-axis if they are set as free parameters. As with monochromatic electrons, the set of accessible states lies close to the ground state for large QE radiative decay, while it encompasses a larger portion of the Bloch sphere for large incoming electron rate, approaching the point of maximally mixed states ($z=0$) for $\gamma_0/\gamma_e\rightarrow0$. The map shows numerical results, as no closed-form expression for the regions could be found. This makes the prediction of final QE states for a particular subset of electron wavefunction parameters a non-trivial task. However, it can be shown that for a given value of $\gamma_0/\gamma_e$, the sphere of radius $r_{\rm s}=\gamma_e/(2\gamma_e+\gamma_0)$ and center $z_{\rm s}=r_{\rm s}-1$ is inscribed within the region of accessible states (and therefore, all states inside it are attainable). Also, a expression for the steady states in the limit of weak coupling strength can be found in~\cite{Zhao2021}. Note that the dashed black line in Figure~\ref{fig:fig2} renders the edge of the Bloch sphere, where QE pure states are located. Thus, only accessible steady states in the vicinity of the ground state have high purity, which imposes an essential limitation for quantum state preparation purposes. 

Once we have investigated the QE steady state, we pay attention next to the transient dynamics of the interaction between emitter and incoming free electrons. The principal timescales of the system are given by the eigenvalues of the coefficient matrix in Equation~\eqref{eq:Master_eq_Bloch}. In the case of monochromatic electrons, it is already diagonal, with all eigenvalues being real and negative: $\lambda_{1,2}=-(\gamma_0 +g_1)$ and $\lambda_3=-(\gamma_0+2 g_1)$. The system dynamics therefore consists of an exponential decay of the Bloch vector to the steady state, which takes place even in the absence of radiative QE decay, through the electron-induced dephasing mechanisms introduced in the previous section, which is weighted by $g_1\propto \sin^2(\beta)$. 

\begin{figure}[t!]
\includegraphics[angle=0,width=\columnwidth]{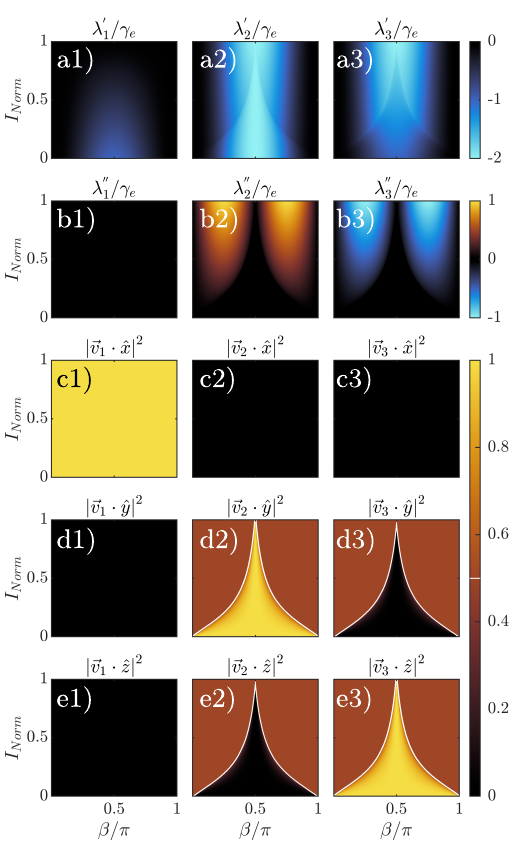}
\caption{Real (a) and imaginary (b) part of the three eigenvalues, $\lambda_{1,2,3}$ , of the coefficient matrix in Equation~\eqref{eq:Master_eq_Bloch} for $I_1=I_2=I_{Norm}$ and $\gamma_0=0$. The eigenvalues are rendered versus the absolute value of the overlapping integral and QE-electron interaction strength. $x$ (c), $y$ (d) and $z$ (e) components of the eigenvectors, $\vec{v}_{1,2,3}$, as a function of $I_{Norm}$ and $\beta/\pi$.} 
\label{figure:fig_eigenvalues_eigenvectors}
\end{figure}

Figure~\ref{figure:fig_eigenvalues_eigenvectors} displays the real (a) and imaginary (b) part of the eigenvalues, $\lambda_{1,2,3}$, for real $I_1=I_2=I_{Norm}$ and $\gamma_0=0$. These are rendered as a function of $I_{Norm}$ and the coupling strength $\beta/\pi$. Note that $\gamma_0$ enters in the coefficient matrix of Equation~\eqref{eq:Master_eq_Bloch} through the identity matrix, which means that its effect is limited to a shift in the real part of the eigenvalues.  All eigenvalues have negative real part and present vanishig (b1), positive (b2) or negative (b3) imaginary part. Thus, all describe an exponential decay into the steady state, but also an oscillatory Rabi-like dynamics, which will give rise to transient states beyond those in Figure~\ref{fig:fig2}. This oscillatory behaviour only takes place in the lateral regions of panels (b2) and (b3), where $\lambda_{2,3}''/\gamma_e$ are non-zero. These are delimited by the inequality $|I_{Norm}||\sin(2\beta)|>\sin^2(\beta)$ (Rabi frequency larger than electron dephasing rate). 

In the limit of weak QE-electron coupling, the eigenvalues of the coefficient matrix acquire the form
\begin{flalign}
&\frac{\lambda_1}{\gamma_e}=-\frac{\gamma_0}{\gamma_e} -{\beta}^2 \left[1-|I_{2}|\,\cos\left(2\,\theta _{1}-\theta _{2}\right)\right], \nonumber \\
&\frac{\lambda_2}{\gamma_e}=-\frac{\gamma_0}{\gamma_e}-\beta ^2\frac{\left(3+|I_{2}|\,\cos\left(2\,\theta _{1}-\theta _{2}\right)\right)}{2}+2{}\mathrm{i}|I_{1}|\beta,\label{lambdas} \\ 
&\frac{\lambda_3}{\gamma_e}=-\frac{\gamma_0}{\gamma_e}-\beta^2 \frac{\left(3+|I_{2}|\,\cos\left(2\,\theta _{1}-\theta _{2}\right)\right)}{2}-2\mathrm{i} |I_{1}|\beta,\nonumber 
\end{flalign}
where we have written $I_n = |I_n|e^{i\theta_n}$. As anticipated above, the imaginary part of the eigenvalues is proportional to the absolute value of $I_1$, while $I_2$ only affects their real part. Thus, we can write the Rabi-like frequency for the system in the weak coupling limit as $\Omega_{R}^{wc}=2|I_1|\gamma_e\beta$~\cite{Zhao2021,Ratzel2021}, which recovers the result in Equation~\eqref{Rabi_H} for $|I_1|=1$. Equations~\eqref{lambdas} reveal that for small $\gamma_0/\gamma_e$ values, the smaller the $\beta$, the more oscillations the QE undergoes before reaching the steady state. They also show that, in these conditions, the phases of the overlapping integrals have an important influence on the exponential relaxation, and therefore the dynamics of the QE state.

Figure~\ref{figure:fig_eigenvalues_eigenvectors} also shows the squared absolute value of the components of the eigenvectors  $\Vec{v}_{1,2,3}$, associated to the three eigenvalues in panels (a)-(b). The components along $x$, $y$, and $z$ are rendered against $I_{Norm}$ and $\beta$ in panels (c), (d), and (e), respectively. Figure~\ref{figure:fig_eigenvalues_eigenvectors}(c) reveals that the projection along $x$ of the Bloch vector representing any QE state is completely governed by $\vec{v}_1$, and therefore experiences an exponential decay in time. On the contrary, the Bloch vector projection within the $yz$-plane does not evolve monotonically, as it involves eigenvectors $\vec{v}_2$ and $\vec{v}_3$. This indicates that the QE excited and ground populations (given by the $z$-component of the state vector)  undergo Rabi oscillations\footnote{In absence of radiative loss, for weak modulation, the z component of the Bloch vector exponentially relaxes to a maximally mixed state and for a QE initially in the ground state one has $z=-e^{\lambda_i^{'} t}$, with an associated excited state probability $P_e(t)=\frac{1+z(t)}{2}$, which at early times $P_e(t)\propto t$. On the contrary, for strong modulation, the complex eigenvalues will appear in complex-conjugate pairs giving $z=-e^{\lambda_2^{'}t}\cos(\lambda_2^{''}t)$, which in the Rabi-dominated regime presents a quadratic excited state probability at early times $P_e(t)\propto t^2$, in agreement with predictions in Ref.~\cite{Ran2022}.}. In fact, the plane of oscillation depends on the phase of the overlapping integrals $I_{1,2}$ (both are real in Figure~\ref{figure:fig_eigenvalues_eigenvectors}). It can be shown that for negligible $|I_2|$, the plane always contains the $z$-axis, and that the angle it forms with $x$ axis is $\varphi=-\atan(1/\tan(\theta_1))=\theta_1-\pi/2$. This behavior is inherited from the ideally modulated case, and stems from the dependence of the QE final populations on the initial coherences. The angle $\psi$ is maximum when $I_1$ and QE coherences are $\pm \pi/2$ out of phase, as illustrated in Equation~\eqref{eq:Probability_gs_ideally_moduated}. 

\begin{figure*}[t!]
\centering
\includegraphics[width=0.95\textwidth]{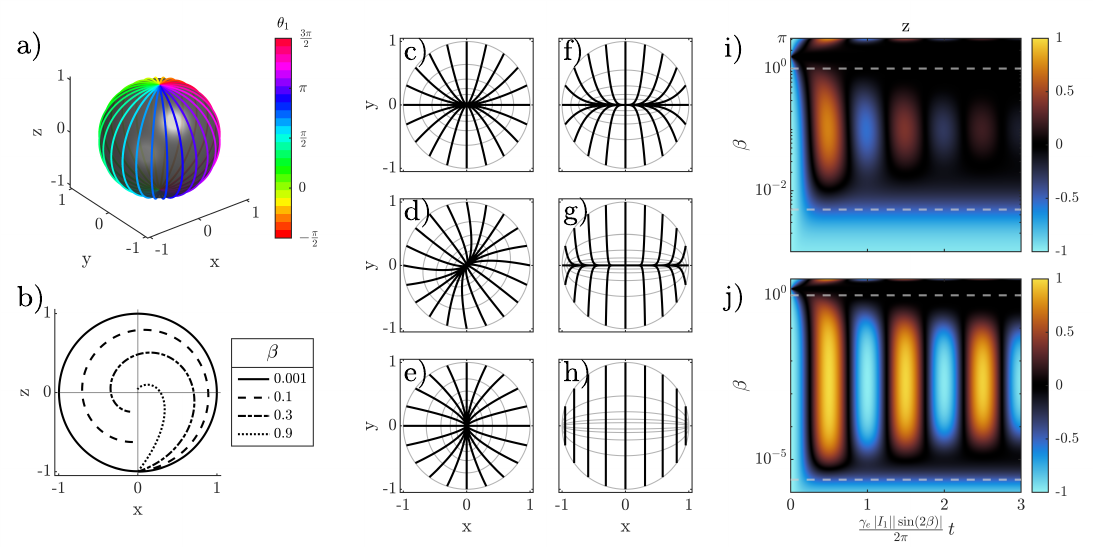}
\caption{QE dynamics under continuous driving with modulated electrons. (a) QE paths along the surface of the Bloch sphere from the ground to the excited state for $\gamma_0=0$ and  different $\theta_1$, the phase of $I_1$. (b) QE trajectories for different $\beta$ values ($\theta_1$ is chosen so that the QE state remains within the $xz$-plane). (c)-(e) Trajectories (black) and equat-time contours (grey) for QEs initially in various pure states within the $xy$-plane. $I_1=0$, $|I_2|=0.2$, and three values of $\theta_2$: $0$ (c), $\pi/2$ (d) and $\pi$ (e). (f)-(h) Same as before but for $\theta_2=0$ and three values of $|I_2|$: $0.5$ (f), $0.9$ (g), and $1$ (h) respectively. (i)-(j) Rabi dynamics undergone by the $z$-component of the Bloch vector as a function of coupling strength for realistic $\gamma_e$ and two QE realizations: solid-state excitons (i) and superconducting qubits (j). Horizontal white lines plot the bounds established by Equation~\eqref{eq:Rabi_dynamics_criteria}.} 
\label{fig:fig3_all_cool_stuff}
\end{figure*}

The dependence of the QE dynamics on $\theta_1$, the phase of the overlapping integral $I_1$, is analyzed in more detail in Figure~\ref{fig:fig3_all_cool_stuff}(a). It renders the trajectories followed by the state vector of a QE initially in its ground state for various values of $\theta_1$. The limit of  negligible purity loss is taken: $\beta=10^{-3}$ and $\gamma_0=0$. We can observe that the QE state lays always at the surface of the Bloch sphere and the trajectory plane rotates with $\theta_1$. This mechanism allows reaching any point on the surface of the Bloch sphere, which shows the capability of modulated free electrons for QE state preparation for low $\beta$ and large $\gamma_e$. The detrimental effect of increasing the QE-electron interaction is showcased in Figure~\eqref{fig:fig3_all_cool_stuff}(b), where $\theta_1=-\pi/2$ and the dynamics take place within the $xz$-plane. The paths followed by the QE state over a single Rabi period, $2\pi/\Omega_{R}^{sc}=\gamma_e|I_1||\sin(2\beta)|$, are displayed for different $\beta$ (note that we have corrected the expression of the Rabi frequency to also account for strong QE-electron coupling). They show the purity loss caused by electron-induced dephasing, which is larger for higher coupling strength, making the QE trajectory depart from the edge of the Bloch sphere and towards the maximally mixed state at $z=0$. 

As illustrated in Equations~\eqref{lambdas}, the overlapping integral $I_2$ allows modifying the decay dynamics of QE state vector. As described in Appendix~\ref{phase_locking}, in general, $I_2$ breaks the symmetric decay of the real and imaginary parts of the QE coherences, and introduces two distinct decay rates for them, $\lambda_{\pm}=-g_1(1\pm |I_2|)$ (note that we take $\gamma_0=0$). The corresponding eigenvectors read
\begin{flalign}
\vec{v}_{\pm}=\frac{\pm\sqrt{1\pm \cos(\theta_2)}\mbox{sign}(\sin(\theta_2))\hat{x}+\sqrt{1\mp \cos(\theta_2)}\hat{y}}{\sqrt{2}}.    
\end{flalign}
In the limit $|I_2|\rightarrow1$, $\lambda_{-}$ vanishes and $\vec{v}_{-}$ is preserved in time, while $\vec{v}_{+}$ decay the fastest. By tuning $\theta_2$, it is possible to tailor the $x$ and $y$-components of $\vec{v}_{-}$, and therefore, to select the phase of the coherences in $\rho^{\rm QE}$. As shown in Appendix~\ref{phase_locking}, the phase preserved fulfils\footnote{If one wishes to preserve the coherences in the plane in which Rabi-oscillations take place, then the phase relationship between the modulation integrals must fulfil $\theta_2=2\theta_1+(2N+1)\pi$, ($N\in \mathbb{Z}$), in agreement with the weak coupling limit eigenvalues in Eq.~\eqref{lambdas}. } $\vartheta \equiv \atan\left(y/x\right)=\theta_2/2+N\pi$. 

The phenomenology outlined above is explored numerically in Figure~\eqref{fig:fig3_all_cool_stuff}(c)-(h), which display QE trajectories (black lines) within the $xy$ plane for various initial pure states (located at the edge of the Bloch sphere) and for different values of $|I_2|$ and $\theta_2$. As a guide for the eye, we show in light grey lines the equal-time contours for initially pure states in the $xy$ plane, as an illustration of their temporal evolution. These clearly demonstrate the asymmetric decay. Again, we take $\beta=10^{-3}$ and $\gamma_0=0$. In the left panels, $|I_2|=0.2$ and $\theta_2=0$ (c), $\pi/2$ (d) and $\pi$ (e), showcasing the rotation of the preserved phase from the horizontal to the vertical direction. Right panels correspond to $\theta_2=0$ and $|I_2|=0.5$ (f), 0.9 (g), and 1 (h). They exhibit a slow dynamics along $\vartheta=0$, and QE trajectories approaching the steady state tangentially to it. Importantly, in Figure~\ref{fig:fig3_all_cool_stuff}(h), a complete preservation of the phase along $\vartheta=0$ is apparent, accompanied by a minimal $\Delta\mathcal{P}$. Note, in contrast, that initial phases orthogonal to it decay into the maximally mixed state. Figure~\ref{fig:fig3_all_cool_stuff}(c)-(h) point towards a strategy to coherently rotate $\rho^{\rm QE}$ within the $xy$-plane by tuning the phase of $I_2$ over time. As detailed in the Appendix~\ref{phase_locking}, by varying the phase of this overlapping integral in time as $\theta_2(t)=2\vartheta_0+\omega t$ (where $\vartheta_0$ is the phase of the coherences in $\rho_0^{\rm QE}$), it is possible to keep the QE's state  phase-locked to $\vartheta$ as long as $\omega \leq 2|I_2|g_1$. 

The main limitation of the quantum state preparation scheme above is that it requires a weak QE-electron coupling. Note that $\beta$ determines the timescale of the coherent dynamics, and therefore, its magnitude relative to the radiative QE lifetime and the electron arrival time set the physical bounds for its implementation. On the one hand, in order to observe Rabi oscillations, the condition $2|I_1|\beta\gamma_e > \gamma_0$ must be fulfilled~\cite{Zhao2021}. On the other hand, for too large coupling, the incoming electrons make the QE state collapse into the maximally mixed state. To avoid this electron-induced dephasing, the condition $2|I_1|\beta\gamma_e > 2\gamma_e\beta^2$ must be met. Both inequalities combined yield
\begin{gather}
    \frac{1}{2|I_1|}\frac{\gamma_0}{\gamma_e} <\beta < |I_1|.  \label{eq:Rabi_dynamics_criteria}
\end{gather}
State-of-the-art electron sources of transmission electron microscopes present repetition rates of up to $40$ MHz~\cite{Abdo2021}, which sets the order of magnitude for $\gamma_e$ above. Depending on the material platform for the QE realization, decay rates many orders of magnitude apart are available. Here we take two as reference: Excitons formed in $\text{WSe}_2/\text{hBN}$ heterostructures, presenting lifetimes of the order of $200$ ns~\cite{Dass2019} ($\gamma_0\simeq4$ MHz), while lifetimes as long as $500$ $\mu$s have been reported in superconducting qubits~\cite{Wang2022} ($\gamma_0\simeq2$ kHz). This means that for perfectly modulated electrons ($|I_1|=1$), the integrated coupling strength, $\beta$, must lay within $\{0.005,1\}$ in the former case, and within $\{2\times10^{-6},1\}$ in the latter. 

Figure~\eqref{fig:fig3_all_cool_stuff} displays the $z$-component of the QE state vector as a function of coupling strength and time for parameters corresponding to the solid-state exciton (i) and superconducting qubit (j). The time axes are normalized to the Rabi period, $2\pi/\Omega_R^{sc}$. The white horizontal lines on each plot correspond to the bounds imposed by Equation~\eqref{eq:Rabi_dynamics_criteria}, which are in excellent agreement with the region of coupling strengths where the oscillatory dynamics takes place. As expected, the oscillations in the QE population are much longer-lived in the superconducting qubit. Note that, in both panels, the QE tends to its ground state ($z=-1$) for $\beta$ below the radiative decay threshold. On the contrary, the QE steady state for $\beta$ above the electron-dephasing threshold is the the maximally mixed state ($z=0$).

\section{\label{section:Quantum_state_tomography} Quantum State Tomography}

So far, we have investigated the exploitation of ideal and realistic (non-ideal) modulated electron wavepackets to perform QE state preparation. In this section, we study the use of the latter as a tool for quantum state readout, which comes as an opportunity that only non-perfectly periodic electron combs offer, as briefly discussed in Section~\ref{section:Ideally_modulated_electrons}. Let us assume that we have prepared an electron in an initial state given as a linear combination of wavevector eigenstates 
\begin{gather}
\ket{\psi_e}= \frac{\sum_{n=-\infty}^{\infty}f_{n}e^{in\phi} \ket{k_0 -n q}}{\sqrt{\sum_{n=-\infty}^{\infty}f_{n}^2}}, \label{eq:free_electron_relative_phase_ch}
\end{gather}
where $f_{n}$ is a real constant and $\phi$ is the phase difference between consecutive wavevector peaks. This electron interacts with a 2-level system, so that the initial state of the system is $\rho_0=\rho_0^{\rm QE}\otimes \dyad{\psi_e}$, and the final state is obtained by the use of the scattering matrix in Equation~\eqref{Magnus} is $\rho=\hat{S}\rho_0\hat{S}^\dagger$. Contrary to previous sections, we trace now over the QE degrees of freedom to obtain the free-electron density matrix after the interaction, $\rho^{e}=\Tr_{\rm QE}\{\rho\}$. This allows us to compute the expectation value for the electron population with wavevector $k$, $\expval{\hat{n}_k}=\Tr\{\hat{c}^\dagger_k\hat{c}_k\rho^e\}$ (with $\hat{c}_k$  being the annihilation operator for an electron of wavevector $k$~\cite{Zhao2021,AbadArredondo2024}), having
\begin{flalign}
    \expval{\hat{n}_k}=&\frac{1}{\sum_{n}f_{n}^2}\sum_n \Big\{\cos(\beta)^2 f_{n}^2 \nonumber\\
    +&\frac{\sin(\beta)^2}{2}\left[f_{n+1}^2+f_{n-1}^2+z\left(f_{n+1}^2-f_{n-1}^2\right) \right] \nonumber \\
    -&\frac{|d|}{2}\sin(2\beta)\sin(\phi_d-\phi)f_{n}\left(f_{n+1}-f_{n-1}\right)
    \Big\}, \delta_{k,k_0-nq} \label{eq:n_k_general_simple_electron}
\end{flalign}
where, for compactness we have defined $d\equiv x+iy$, and $\phi_d\equiv \arg(d)$. Equation~\eqref{eq:n_k_general_simple_electron} reveals that the final electron wavevector distribution is composed by peaks distributed integer multiples away from $k_0$ (the central wavevector of the initial electron wavefunction), and that the final distribution is sensitive to the three degrees of freedom of $\rho_{\rm QE}$~\cite{Ruimy2021,Zhang2022}. Importantly, this indicates that, in principle, it is possible to perform quantum state reconstruction of QE states by modulated free electrons. 

The amount of information carried by the free-lectron wavefunction after the interaction can be measured through its degree of entanglement with the QE state. As we are dealing with a bipartite system in a pure state, we can use the Wooters concurrence~\cite{Rungta2001} as the witness of the electron-QE degree of entanglement
\begin{gather}
    \mathcal{C}\equiv\sqrt{2(1-\mbox{Tr}\{(\rho^e)^2\}}=\sqrt{1-\mathcal{P}},\label{concurrence}
\end{gather}
where $\mathcal{P}$ is the purity of the free-electron reduced density matrix, defined as in Equation~\eqref{eq:Purity definition}. $\mathcal{C}$ ranges from 0, for a maximally mixed state, to 1, for a maximally entangled one. Equation~\eqref{concurrence} establishes that the concurrence is uniquely determined by $\mathcal{P}$, depending monotonically on it, and vanishing as it approaches unity. For the initial electron wavefunction in Equation~\eqref{eq:free_electron_relative_phase_ch} interacting with a QE with an arbitrary density matrix $\rho_0^{\rm QE} = (\mathbb{1}+x\hat{\sigma}_1+y\hat{\sigma}_2+z\hat{\sigma}_3)/2$, we have
\begin{flalign}
\frac{\mathcal{C}^2(\rho)}{\sin(\beta)^2}=
&2(2-|d|^2)(1-{\tilde{I_1} }^2)\,{\cos \left(\beta \right)}^2\nonumber\\
+&(1-z^2)(1- {\tilde{I_2} }^2)\sin(\beta)^2\nonumber\\
+&2 z|d|\sin( \phi-\phi_d)\sin(2\beta)\tilde{I_1}(\tilde{I_2}-1)\\
+&2({\tilde{I_1} }^2-\tilde{I_2})|d|^2 \cos(2(\phi-\phi_d)) \cos(\beta)^2,
\end{flalign}
where we have used that $\sqrt{|d|^2+z^2}=1$ (QE initially in a pure state). If the initial electronic wavefunction is formed by $N$ equal amplitude peaks, then the two overlap integrals behave as $\tilde{I_1}=\max\left\{(N-1)/N,0\right\}$ and $\tilde{I_2}=\max\left\{(N-2)/N,0\right\}$. From these expressions the limiting forms of the concurrence for monochromatic and perfectly modulated electrons can be obtained. These are
\begin{flalign}
\mathcal{C}(\rho)&\overset{N=1}{=}\sin(\beta)\sqrt{4-2|d|^2-(3+z^2-2|d|^2)\sin(\beta)^2},\\
\mathcal{C}(\rho)&\overset{N\rightarrow \infty}{=}0.
\end{flalign}
For an electron wavefunction consisting in a single wavevector peak, the final entanglement depends both on the initial state of the QE and the coupling parameters. The concurrence is maximum for maximum expectation value of the dipole operator ($z=0$) and coupling strength $\beta=\pi/2$. In the case of a perfectly modulated electron, the final concurrence tends to zero. This limiting behavior was already discussed in Section~\ref{sec:stateprep}, where we discarded perfectly modulated electrons as tools for QE quantum state reconstruction. 

\begin{figure*}[]
\centering
\includegraphics[width=\textwidth]{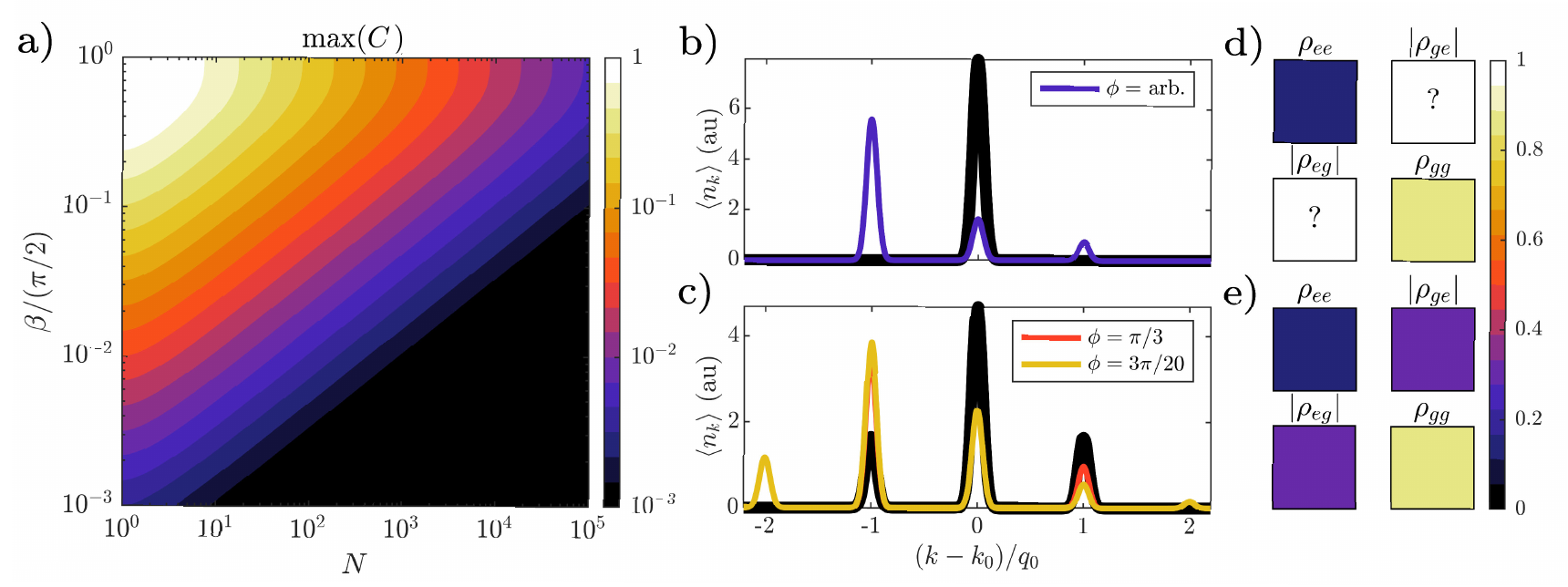}
\caption{Quantum state tomography. \textbf{a)} electron-QE entanglement as measured by the concurrence. \textbf{b-c)} Electronic wavevector distribution before (black) and after interaction with the QE in the case of a monochromatic electron and non-ideal modulated electron. \textbf{d-e)} Quantum state reconstruction from the data obtained in the experiments shown in panels (b,c) respectively, displaying complete state reconstruction for the non-ideal modulated electron.} 
\label{fig:fig4}
\end{figure*}

In Figure~\ref{fig:fig4}(a), we display the maximum concurrence attainable for each free-electron configuration, labelled through the pair of parameters $\{N,\beta\}$ and computed over all possible combinations of $\rho_0{^{\rm QE}}$ and electronic modulation phase, $\phi$. It shows that the degree of entanglement monotonically diminishes with decreasing coupling strength and comb size. High concurrence is possible for $N>1$, which indicates that, in principle, information can be extracted efficiently from the system, while having access to the phase properties of the QE coherences. In what follows, we explore this idea of using electron combs with few modulation wavevector peaks to readout the state of a QE, and compare this strategy with using monochromatic wavefunctions ($N=1$).

We analyze first the monochromatic wavefunction by feeding Equation~\eqref{eq:n_k_general_simple_electron} with $f_n=\delta_{n0}$. The final electronic state is composed by three wavevector components: zero-loss peak, energy loss and energy gain sidebands. Their amplitudes encode the probabilities for the electron wavevector to remain as before the interaction, or to gain, or lose a quanta matching the QE optical transition. These are given by
\begin{flalign}
    \expval{\hat{n}_{k_0}}&=\cos(\beta)^2, \label{eq0}\\
    \expval{\hat{n}_{k_0\pm q}}&=\frac{\sin(\beta)^2}{2}\left[1\pm z\right].\label{eqpm}
\end{flalign}
Equation~\eqref{eq0} makes evident that the zero-loss peak depletion gives information about the QE-electron coupling strength, while the imbalance between energy gain and loss in Equation~\eqref{eqpm} reveals the presence of an initial population in the QE excited state. Thus, we conclude that no information about the QE coherences can be extracted from the free-electron wavefunction, and therefore it is impossible to ascertain the quantum nature of the emitter state. Pure and mixed states sharing the same Bloch $z$-component would lead to exactly the same electron wavevector spectra. This we show in Figure~\ref{fig:fig4}, where panel (b) plots the spectrum for a monochromatic wavefunction before (black) and after (purple) the interaction with a QE prepared in a certain initial state. One can see how the zero-loss peak is depleted by the interaction with the QE and the energy loss and gain sidebands appear with different amplitudes. From this spectrum, the QE state reconstruction in Figure~\ref{fig:fig4}(d) was obtained. The lack of information about the coherences (represented by question marks) makes the assessment of the purity of the QE state impossible.

Equation~\eqref{eq:n_k_general_simple_electron} shows that at least 2 wavevector components spaced by $q$ are needed in the initial electron wavefunction to probe the QE coherences~\cite{Ruimy2021}. We consider an electron wavepacket with initial wavevector components $\ket{k_0}$ and $\ket{k_0\pm q}$, with $f_1=f_{-1}$. After the interaction with the QE, 5 different wavevector components emerge in the final electron state, with spectral amplitudes given by
\begin{flalign}
    \expval{\hat{n}_{k_0}}=&\cos(\beta)^2 \alpha_0^2
    +\frac{\sin(\beta)^2}{2} \alpha_1^2 \\
    \expval{\hat{n}_{k_0\pm q}}=&\frac{\cos(\beta)^2}{2} \alpha_1^2 +\frac{\sin(\beta)^2}{2}\left[1\pm z\right] \alpha_0^2, \nonumber\\ \label{q}
    \mp&\frac{|d|}{2\sqrt{2}}\sin(2\beta)\sin(\phi_d-\phi)\alpha_0\alpha_1,\\
    \expval{\hat{n}_{k_0 \pm 2q}}=& \frac{\sin(\beta)^2}{4}\alpha_1^2\left[1\pm z\right],\label{2q} 
\end{flalign}
where $\alpha_1=\sqrt{2} f_1/\sqrt{f_0^2+2f_1^2}$ and $\alpha_0=  f_0/\sqrt{f_0^2+2f_1^2}$. We introduce the symmetric and antisymmetric combination of the $n$-th spectral amplitude as $n_j^S\equiv \expval{\hat{n}_{k_0 +jq}}+\expval{\hat{n}_{k_0 -jq}}$ and $n_j^A\equiv \expval{\hat{n}_{k_0 +jq}}-\expval{\hat{n}_{k_0 -jq}}$. These can be evaluated the Equations~\eqref{q} and \eqref{2q}, and the magnitudes characterizing the initial state of the system can be obtained as 
\begin{flalign}
    z&=\frac{n_2^A}{n_2^S},\label{eq:z_tomography}\\
    \sin(\beta)^2&=\frac{2n_2^S}{\alpha_1^2}, \label{eq:beta_tomography}\\
    |d|\sin(2\beta)\sin(\phi-\phi_d)&=\frac{\sqrt{2}}{\alpha_0 \alpha_1}\left(n_1^A-2\frac{\alpha_0^2}{\alpha_1^2} n_2^A\right).
\end{flalign}

The initial QE populations are uniquely specified by Equation~\eqref{eq:z_tomography}, while it is apparent that the coupling strength is not completely determined by Equation~\eqref{eq:beta_tomography}, as it can have any of 4 different values: $\beta=\left\{\beta_0,\pi-\beta_0,\pi+\beta_0,2\pi-\beta_0\right\}$, with $\beta_0=\sin^{-1}(\sqrt{2 n_2^S}/|\alpha_1|)$. For weak electron-QE interaction, the only relevant option is $\beta=\beta_0$. To extract the QE coherences, one can perform the measurement with different $\phi$ (in the initial modulated wavefunction). It can be shown that two different phases are sufficient to determine coherences of the system. By defining $\gamma_j=\frac{\sqrt{2}}{\alpha_0 \alpha_1}\left(n_1^A-2\frac{\alpha_0^2}{\alpha_1^2} n_2^A\right)$ as the estimation corresponding to the phase $\phi_j$, the coherences are given by
\begin{flalign}
d= \frac{1}{\sin(2\beta_0)} \frac{ \left[ \gamma_1 e^{i \phi_2}-\gamma_2 e^{i\phi_1} \right]}{\sin(\phi_1-\phi_2)},
\end{flalign}
which shows that complete arbitrary QE state retrieval is possible under modulated electron probing. We showcase this protocol in Figure~\ref{fig:fig4}(c), where the initial (black) and final (orange, yellow) wavevector spectra for a QE in a certain state are displayed. Note that the free-electron spectra after the interaction are different for the two modulation phases, $\phi_{1,2}$. This proves that the initial QE state has non-zero coherences, and it is therefore not a statistical mixture. By applying the strategy outlined above, we can construct the initial QE density matrix, as shown in Figure~\ref{fig:fig4}(f) where, in contrast to Figure~\ref{fig:fig4}(d), a full characterization of the QE coherences is possible.

Let us clarify that the QE coherences in Figure~\ref{fig:fig4}(f)  were obtained under the assumption that the interaction strength is positive and real. For general complex $\beta$, we can follow the procedure described in Section~\ref{section:Ideally_modulated_electrons}: by performing a basis change and absorbing the phase of the coupling strength into the ground state of the QE, the scattering matrix takes the same form as if the coupling strengths were real and positive. This means that it is enough to multiply the QE coherence in Figure~\ref{fig:fig4}(f) by $\exp{-i\phi_\beta}$ to extend the result to complex coupling strengths, while the $z$-component of the Bloch vector remains unchanged. Therefore, in general, to perform proper state readout it is needed to have characterized the complex QE-electron coupling strength in a previous experiment. This can be done by extending recently proposed, homodyne-based ideas in the realm of bosonic modes~\cite{Gaida2024} to QEs. Note that that for interactions with an isolated QE, this phase factor is irrelevant, while for ensembles of interacting QEs, the relative phases of their ground states can have an important role. Finally, we remark that, even in the situation in which the coupling strength is complex valued, our protocol allows determining unequivocally the purity of the initial QE state, as $\mathcal{P}=\sqrt{z^2+ |d|^2}$, is insensitive to phase shifts in $d$. This is a direct consequence  of the coherent interference effects taking place in modulated wavefunctions, which are absent in monochromatic electrons. 

\section{Conclusion}

We have explored the quantum interaction of modulated free-electrons and single quantum emitters. First, we have considered single electron-emitter interaction events, and revealed that the relevant figure of merit for the electron wavefunction is the autocorrelation function in reciprocal space. In the limit perfectly periodic (ideal) modulation, we have shown that the purity of the emitter state is preserved, and emitter and free electrons remain disentangled after their interaction. Moving into a configuration of continuous electron driving, we have revealed Rabi-like dynamics in the emitter state, and explored its interplay with two decoherent effects: radiative losses and electron-induced dephasing. A previously unexplored mechanism of phase-locking has been reported, that takes place upon continuously changing the phase of the electron modulation. Finally, we have focused on quantum state tomography, and by analyzing the degree of entanglement between emitter and free electron, we have determined that small combs are best suited for the task. Thus, a measurement protocol able of uniquely determining the purity of the emitter state has been proposed. Our theoretical findings prove that, beyond their extremely high spatial resolution, modulated electrons provide multiple opportunities to precisely prepare and fully characterize the quantum state of single quantum emitters. We believe this makes them a promising tool for quantum technologies, where their power and versatility can be exploited to address and manipulate qubits one by one.

\begin{acknowledgments}
We thank F.J. García-Vidal for useful discussions. We acknowledge financial support from the Proyecto Sinérgico CAM 2020 Y2020/TCS-6545 (NanoQuCo-CM), and MCINN projects PID2021-126964OB-I00 (QENIGMA) and TED2021-130552B-C21 (ADIQUNANO).
\end{acknowledgments}


\appendix

\section{Magnus expansion} \label{sec:Magnus_expansion_first_2_terms}
Given a first order differential equation $d\ket{\psi(t)}/dt=\hat{A}(t)\ket{\psi(t)}$, the solution for the initial value problem is in general given by the time ordered operator presented in \cref{eq:TimeOrdering_scatteringmatrix}. However, one can make use of the Magnus expansion to construct the propagator as the true exponential of a matrix as 
\begin{gather*}
    \ket{\psi(t)}=e^{\hat{\Omega}(t)}\ket{\psi(t=0)},
\end{gather*}
where the $\hat{\Omega}$ operator accepts a series decomposition $\hat{\Omega}(t)=\sum_n\hat{\Omega}_n(t)$, where the first two terms in the expansion read \cite{Blanes2010}
\begin{flalign}
\hat{\Omega}_1(t)&=\int_{0}^t \hat{A}(t_1) dt_1,\\
\hat{\Omega}_2(t)&=\frac{1}{2}\int_{0}^{t} dt_1 \int_{0}^{t_1} d t_2 [\hat{A}(t_1),\hat{A}(t_2)].
\end{flalign}
In our case, the $\hat{A}$ operator corresponds to the interaction Hamiltonian in the interaction picture as $\hat{A}(t)\equiv -i \hat{H}_{I,{\rm int}}(t) /\hbar$. For the interaction Hamiltonian shown on \cref{eq:IntHam0}, it can be shown that, under the non-recoil approximation,  
\begin{gather}
\hat{A}(t)=-i \sum_q \left[g_q \hat{b}_q \hat{\sigma}^\dagger e^{i\left(\frac{\omega^*}{v_0}- q\right)v_0t} +g_q^* \hat{b}_q^\dagger \hat{\sigma} e^{-i\left(\frac{\omega}{v_0}-q\right)v_0t}\right].
\end{gather}
Here, we give the first two terms of the Magnus expansion as $t\rightarrow \infty$. Note that $\omega=\omega'-i\gamma$ is the QE frequency, where $\gamma$ is a small, positive, imaginary part that ensures the QE relaxation for $t\rightarrow \infty$, and which will disappear in the end of the calculation once the limit $\gamma\rightarrow 0$ is taken. As short-hand notation we introduce $\phi_q\equiv \omega-v_0 q$, and $\phi^*_q\equiv\omega^*-v_0 q$ for the phases accumulated by the different terms in the interaction Hamiltonian. The relevant conmutator for the calculation is
    \begin{flalign}
        [A(t_1),A(t_2)]=& \nonumber\\
        =\sum_{q,q'} g_q^* g_{q'}\hat{\sigma}_z & \hat{b}_{q'-q}  \left[e^{i \phi^*_{q'} t_2}e^{-i\phi_{q} t_1}-e^{i \phi^*_{q'} t_1}e^{-i\phi_{q} t_2} \right],
    \end{flalign}
The first two terms of the Magnus expansion are given by
\begin{flalign}
    &\hat{\Omega}_1(t)\overset{t\rightarrow\infty}{=}\hat{\Omega}_1=-i \left[\beta \hat{b}_{q_0} \hat{\sigma}^\dagger +\beta^* \hat{b}_{q_0}^\dagger \hat{\sigma}\right],\\
    &\hat{\Omega}_2(t)\overset{t\rightarrow\infty}{=}\frac{i}{2}\frac{|\beta|^2}{2\pi}\hat{\sigma}_z \int dq   \Bigg[ \frac{g_{q}g_{q_0}^{*} \, \hat{b}_{q-q_0}}{q-q_0 -i\gamma/v_0} +h.c.\Bigg]=0,
\end{flalign}
where $q_0=\omega'/v_0$ is the wavevector lost/gained by the incoming electron after exchanging one resonant photon with the QE in the non-recoil approximation. The first term captures all the dynamics present in this work~\cite{Ruimy2021}. The vanishing of the second one ensures that the results in this work have an associated error of $O(\beta^{3})$.

\section{overlapping integrals for PINEM electrons after drift} \label{sec:Mod_integrals_Gover}
From Ref.~\cite{Zhang2022}, the wavefunction of a free electron after a PINEM preparation step and a drift-time $t_D$ is given by $\ket{\psi}=\int dk B(k) \ket{k}$ with
\begin{flalign}
B(k)=\sum_m e^{-im\phi_0} \int dk \frac{e^{-i E_{\hbar k} t_D/\hbar}}{(2\pi \sigma_k^2)^{1/4}}
J_m(2|g_L|) e^{-\frac{(k-k_0-m q_0)^2}{4\sigma_k^2}},
\end{flalign}
where $\sigma_k$ is the initial wavevector spread of the free electron, assumed to be much smaller than the recoil $\sigma_k\ll q_0$, $g_L$ is the interaction strength with the PINEM field, $k_0$ is the initial wavevector of the free electron, and $g_L$ is the integrated coupling strength, defined as in Section~\ref{sec:theory}. We consider an  energy dispersion relation for the free electron with relativistic corrections given by $E_p=E_0+v_0 (p-p_0)+\frac{1}{2\gamma^3m_e}(p-p_0)^2$. Therefore, with this wavefunction, the overlapping integrals $I_n=\expval{\hat{b}^n}$, have the form
\begin{flalign}
    I_n=&\exp\left[ -i n (\phi_0+ \hbar q_0 v_0 t_D)-i \frac{n^2}{2} \theta  -\left( \frac{n \sigma_k}{\sqrt{2} q_0} \theta \right)^2 \right] \times \nonumber \\
    &\times \sum_{m} J_m(2|g_L|)J_{m+n}(2|g_L|) e^{-i\,  m\, n\, \theta}.
\end{flalign}
We have introduced a new definition for the phase factor appearing in the summation as 
\begin{flalign}
    \theta=\left( \frac{q_0 v_0}{\omega} \right)^2  \omega\, t_D \left( \frac{1-\eta^2}{\eta^2}\right)\frac{ \hbar \omega }{\gamma m_e c^2 },
\end{flalign}
where $\eta = v_0 /c$ and we have also made $q_0=\omega/v_0$. Note that since $\sum_m J_m(x)J_{m+n}(x)=\delta_{n,0}$, the fact that this phase factor acquires non-zero values is key to achieve a significant degree of modulation through the drifting process. Also note that $\theta$ also 
appears as a damping term within the gaussian, which implies that when electrons have vanishing speeds, since $\theta\rightarrow\infty$, all the overlapping integrals are exponentially attenuated, just like predicted by the non-relativistic treatment. On the other hand, for electron speeds close to the speed of light, $\theta \rightarrow 0$, and the overlapping integrals vanish. This illustrates how a careful balance between electron speed and drift time must be attained. 

In the case of the first overlapping integral, we have   
\begin{gather*}
    I_1 \propto \sum_{m} J_m(z)J_{m+1}(z) e^{-i\,  m\, \theta}=\\
    =-\frac{1}{2} \frac{\partial}{\partial z} \left[ \sum_{m=-\infty}^\infty J_m^2(z) e^{-i\,  m\,\theta} \right]-\frac{2i}{z}\sum_{m=1}^\infty m J_m(z)^2 \sin(m\theta)
\end{gather*}
where we have used the recurrence relation of the bessel functions~\cite{Abramowitz1988}. Although general solution of these sums couldn't be obtained, one can get analytical expressions for particular values of $\theta$ that will prove useful. Two of such cases correspond to $\theta$ being be an even or odd integer multiple of $\pi$, which yields
\begin{gather*}
I_1\propto
    \begin{cases}
			0, & \text{if $\theta=2N\pi$}, \\
            J_1(4|g_L|), & \text{if $\theta=(2N+1)\pi$},
		 \end{cases}
\end{gather*}
with $N\in \mathbb{Z}$. Throughout this paper we have shown that the closer the overlapping integrals are to unity, the closer the modulated electrons are to perfectly periodic electron combs. Therefore it is of interest to determine the upper bound of the overlapping integrals that can be attained in this way. We have numerically investigated this matter, and found that $\theta=\pi$ establishes a global upper bound for attainable modulation and therefore
\begin{gather}
|I_1|\leq \exp\left[-\left( \frac{\pi \sigma_k}{\sqrt{2} q_0}\right)^2 \right] J_1(4|g_L|).
\end{gather}
The Bessel function acquires it's maximum value at $4|g_L|=0.46$, which in the limit of $\sigma_k\rightarrow 0$, gives that the first overlapping integral for an electron prepared with a PINEM experiment followed by a drift protocol can never surpass $|I_1|\lessapprox 0.581$, which is far from the required values for purity preservation during the interaction with QEs.

\section{Phase locking with non-ideal modulated electrons}\label{phase_locking}
Equation~\eqref{eq:Master_eq_Bloch} describes the evolution of the QE state vector in the Bloch sphere. By neglecting $I_1$ and $\gamma_0$, so as to focus on the phase locking behavior described in Section~\ref{sec:stateprep}, the $z$-component becomes decoupled and we have
\begin{gather}
\begin{pmatrix}
\dot{x} \\
\dot{y} 
\end{pmatrix} = \begin{pmatrix}
-g_1\left(1-I_2^{'}\right)& g_1 I_2^{''} \\
g_1 I_2^{''}& -g_1\left(1+I_2^{'}\right)
\end{pmatrix}
\begin{pmatrix}
x \\
y 
\end{pmatrix} 
\end{gather}
By writing $x$ ($y$) in polar coordinates, $x=r\cos(\vartheta)$ ($y=r\sin(\vartheta)$) we find that the  EOM can be written as
\begin{flalign}
    \frac{\partial_\tau{r}}{r}=&-1+|I_2|\cos(\theta_2-2\vartheta)\label{eq:radial_change_phase_lok},\\
    \partial_\tau{\theta}=&|I_2|\sin(\theta_2-2\vartheta),\label{eq:angle_change_phase_lok}
\end{flalign}
where we have written $I_2=|I_2|\exp(i\theta_2)$, and the time derivatives are with respect to $\tau = g_1 t$. Equations~\eqref{eq:radial_change_phase_lok} and \eqref{eq:angle_change_phase_lok} reveal fixed points at $\vartheta=(n\pi+\theta_2)/2$, where $n$ is an integer. Plugging these angles into Equation~\eqref{eq:radial_change_phase_lok}, we see that whenever $n$ is even (odd) the radial decay is slowed (accelerated), and in the limit of $|I_2| \rightarrow 1$ the decay rate becomes zero. By linearizing Equation~\eqref{eq:angle_change_phase_lok} around these fixed points, we see that whenever $n$ is even (odd) the fixed point is stable (unstable). Thus, this interaction preserves the phases equal to $\theta_2/2$ or $\theta_2/2 +\pi$. We now make $\theta_2$ time dependent, changing linearly in time, so that $\theta_2=\omega t$. This implies that the equations of motion can be casted as
\begin{flalign}
    &\frac{\dot{r}}{r}=\left(-1+|I_2|\cos(\theta_2-2\vartheta)\right)g_1, \nonumber \\
    &\dot{\vartheta}=g_1|I_2|\sin(\theta_2-2\vartheta),\nonumber \\
    &\dot{\theta}_2=\omega. \nonumber 
\end{flalign}
We now define a new variable, the phase difference as $\Delta\vartheta=\theta_2-2\vartheta$, which allows to write 
\begin{flalign}
    \frac{\dot{r}}{r}=&\left(-1+|I_2|\cos(\Delta\vartheta)\right)g_1, \label{eq1}\\
    \dot{\Delta\vartheta}=&\omega -2g_1|I_2|\sin(\Delta\vartheta),\label{eq2}
\end{flalign}
which shows that whenever $\kappa\equiv \omega/(2g_1|I_2|)\leq 1$, phase locking is possible between the impinging electron wavefunction and the QE state. Note that the fixed point in phase difference corresponds to $\Delta\vartheta_0=\asin(\kappa)$. 

By writing $\Delta\vartheta=\Delta\vartheta_0 + \delta$ and linearizing Equations~\eqref{eq1} and \eqref{eq2} around the fixed point, we find that ${\dot{r}=r\left(-1+|I_2|\left[\sqrt{1-\kappa^2}-\kappa \delta(t)\right]\right)g_1}$ and ${\delta(t)=\delta_0 \exp\left( -2g_1|I_2|\sqrt{1-\kappa^2} t\right)}$. Therefore, as long as $\kappa < 1$, the initial phase difference will tend to shrink to the one set by the fixed point. In the limit of $\kappa\rightarrow 0$, $|I_2|\rightarrow 1$ the QE vector state components become
\begin{flalign}
    r(t)=&r_0 \exp\Big[-\frac{g_1\kappa^2}{2}\left(t+\frac{\delta_0}{4\kappa g_1 |I_2|}\left( 1-e^{-2g_1|I_2|t}\right) \right) \Big],\\
    \delta(t)=& \delta_0 \exp\left( -2g_1|I_2|t\right),\\
    \theta(t)=& \frac{\omega}{2} t -\frac{\kappa}{2}  + \frac{\theta_2(0)-\left[\theta_2(0) -2\vartheta(0)\right]e^{-2g_1|I_2|t}}{2}. 
\end{flalign}
We can see that in the transient in which the QE state reaches the phase locking, its purity is scaled by a factor of $\exp(-\delta_0 \kappa/4|I_2|)$, and then it slowly decays as $\exp(-g_1\kappa^2 t)$. The steady state is phase locked with $I_2$, and the initial difference between the two phases vanish over time. This gives access to another way of inducing rotations in the Bloch sphere. 


\bibliography{bib_electrons}
\end{document}